\documentclass[preprint,12pt]{elsarticle}
\usepackage{graphics,bm}
\usepackage{amssymb}
\usepackage{epsfig}
\usepackage{epsf}
\usepackage{graphicx}
\usepackage{mathrsfs}
\usepackage{amsfonts}
\makeatletter

    \usepackage{amssymb}
    \usepackage{amsmath}
    \usepackage{amsbsy}
    \usepackage{bm}
    \usepackage{graphicx}
 \usepackage{xcolor}
 \usepackage{calrsfs}
\DeclareMathAlphabet{\pazocal}{OMS}{zplm}{m}{n}

\newcommand{\Vb}{\pazocal{V}}

\newcommand{\Rb}{\pazocal{R}}
\newcommand{\Eb}{\pazocal{E}}

\newcommand{\Ab}{\pazocal{A}}

\journal{Annals of physics}
  
\begin{document}
\begin{frontmatter}

\title{Spin wave vortex from the scattering on Bloch point solitons}

\author[rvt,focal]{V. L. Carvalho-Santos}
\ead{vagson.carvalho@usach.cl}

\author[focal]{R. G. El\'ias}
\ead{gabriel.elias@usach.cl}

\author[saya]{A. S. N\'u\~nez}
\ead{alnunez@dfi.uchile.cl}

\address[rvt]{Instituto Federal de Educa\c c\~ao, Ci\^encia e Tecnologia
Baiano - Campus
Senhor do Bonfim, \\Km 04 Estrada da Igara, 48970-000 Senhor do
Bonfim, Bahia, Brazil.}
\address[focal]{Departamento de F\'isica, Universidad de Santiago de Chile and CEDENNA, Avda. Ecuador 3493, Santiago, Chile.}
\address[saya]{Departamento de F\'isica, Facultad de Ciencias F\'isicas y 
Matem\'aticas, Universidad de Chile, Casilla 487-3, Santiago, Chile}

\begin{abstract}
The interaction of a spin wave with a stationary Bloch point is studied. The topological non-trivial structure of the Bloch point manifests in the propagation of spin waves endowing them with a gauge potential that resembles the one associated with the interaction of a magnetic monopole and an electron. By pursuing this analogy, we are led to the conclusion that the scattering of spin waves and Bloch points is accompanied by the creation of a magnon vortex. Interference between such a vortex and a plane wave leads to dislocations in the interference pattern that can be measurable by means of magnon holography.   
\end{abstract}


\end{frontmatter}

%+++++++++++++++++++++++++++++++++++++++++++++++++++++++++++++++++++

\section{Introduction}
In a magnetic material textures in the local magnetic ordering spread across in the form of spin waves (SWs). Such disturbances correspond to elementary excitations with zero associated charge and net spin equals to unity. The field of solid state physics concerning the manipulation, detection and dynamics of the SWs in a magnetic system \cite{Kruglyak2010} has been dubbed magnonics.  The field of magnonics has grown into a well established realm of magnetism and opened new paths in the understanding of magnetization dynamics of complex structures.
 The wide frequency range that magnonic excitations can display (from GHz to THz \cite{Cherepanov199381}), the possibility of tailoring the SW spectrum in the so-called magnonic crystals \cite{Krawczyk-2014},  the possibility of manipulating magnetic textures by SWs, \cite{Han2009} the interference pattern of SWs in the presence of geometrical constrictions or of magnetic solitons, \cite{Hertel-2004fg} and the observation of the particular geometry of the scattered SWs after passing through magnetic solitons \cite{Elias-2014gf} are simple examples of the rich tapestry of  phenomena in the fields. 
 Interestingly, quantum effects can play a crucial role in the physics  of macroscopic assemblies of magnons, leading to the collective behaviour in the form of Bose-Einstein condensates even at room temperature. \cite{byrnes-2014aa,serga-2014aa,Troncoso-2014,Troncoso-2012} Such a variety of phenomena has attracted interest from a diversity of sources regarding its potential applications in technological SW devices. \cite{Xing-2013,Khitun-2013,Csaba-2014,chumak-2015aa} The different contributions of the magnetic energy play a different role depending on the scales associated with the SW phenomena.
It is from this perspective that the SW wavelength $\lambda$ can be used to separate the behaviour into two regimes. \cite{a.-g.-gurevich-1996aa} Short wavelengths ($\lambda <1\; \mu m$) are dominated by the short range exchange interaction. This kind to SWs are named exchange SWs. On the other hand, in the behaviour of SWs of longer wavelengths the physics is dominated by the long range dipolar interaction, in this regime we talk about dipolar or magnetostatic SWs. In this paper we will study SW phenomena of short wavelength and the physical behaviour will be dominated by the exchange interaction. Our basic model will, therefore, start with an analysis of the SW phenomena in presence only of the exchange interaction, leaving all the effects arising from other energy contributions in the form of perturbations.

Our main focus will be in the interplay between SWs and a background magnetic texture. This effect has been studied in several papers in the context of SW propagation across domain walls, vortices and skyrmions. \cite{Gonzalez-2010aa,Park-2005aa, Dugaev-2005aa,Ivanov-2007,Iwasaki-2014} A smooth texture affects the dynamics of a SW by changing the basic wave equation that rules the SW behaviour. A very interesting idea has been put forward by Dugaev et al. \cite{Dugaev-2005aa}  that showed that the magnetic texture induces Berry phases \cite{berry-1984aa} and Berry-curvatures and applied these ideas to study the Aharanov-Bohm-like \cite{Aharanov-1959} effect of the magnetization in a magnetic ring with a non-trivial magnetic configuration.  
Our analysis start from the exchange Hamiltonian and derive an effective Hamiltonian for the SWs that is equivalent to a $O(2)$ gauge theory. The role of the Berry curvature predicted by Dugaev et al. enters in our theory in the form a $O(2)$ vector potential. In addition to the Berry connection induced by the background spin texture we find an additional term that resembles a scalar potential. From this perspective we derive a Noether current that is conserved. We study under what circumstances the behaviour of the effective theory can be cast in the form of a Schr\"odinger-like equation for the magnon amplitude. In this way we can easily derive the corresponding effective magnetic field felt by the SWs because of the texture. 
After a revision of the insights gained putting SWs in a gauge-invariant form, we focus on the particular texture given by a three dimensional soliton, with non zero topological charge, known as Bloch point (BP). \cite{Doring-1968bf,Elias-2011fk} After evaluating the different potentials, we conclude that the basic physics of the SWs in presence of a BP corresponds to an effective theory of an electron in presence of a magnetic monopole. Our main result is that in the context of the interaction of SWs with a BP the outgoing pattern correspond to a SW vortex. This interesting result can be studied from an experimental point of view by using the technique of magnon-holography \cite{Khitun-2013}. \\

\begin{figure}
\centering
\includegraphics[scale=0.36]{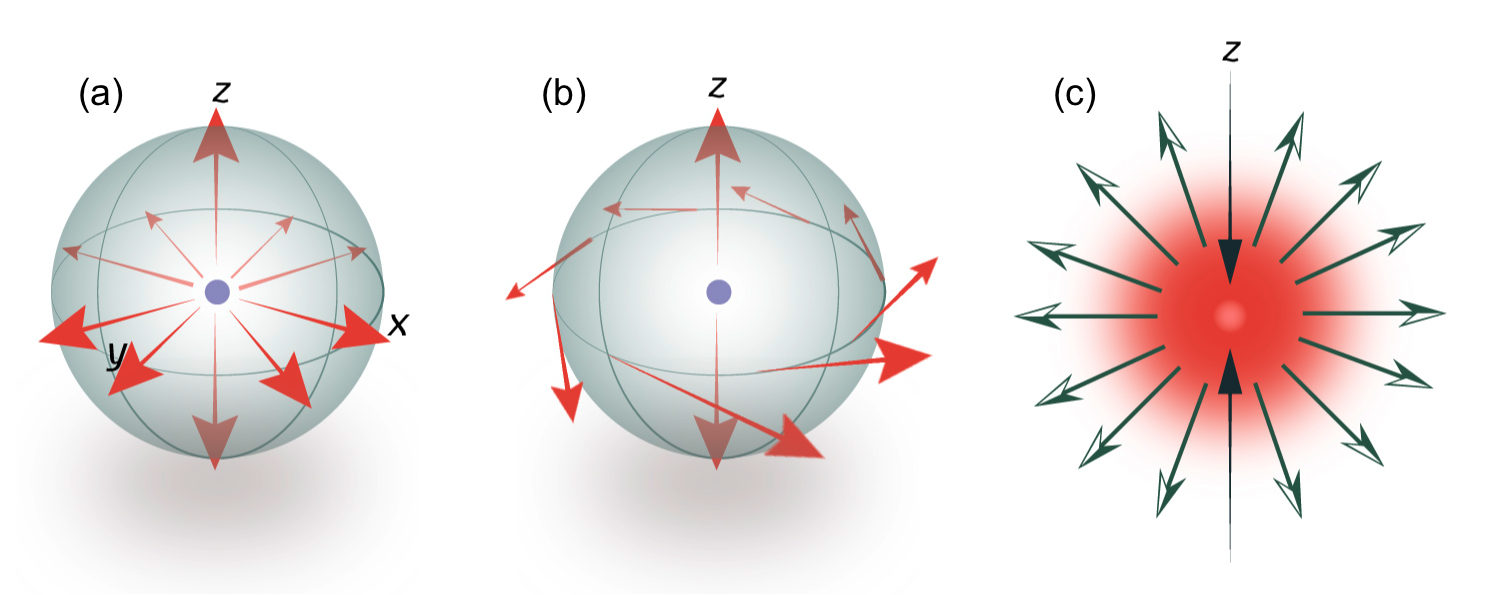}
\caption{Different Bloch point singularities. (a) Bloch point with $\gamma=0$, (b) Bloch point with $\gamma=\pi/2$  and (c) effective magnetic field seen by the SWs. We have highlighted the singularities associated with the Dirac strings along the $\pm \bm z$-axis.}
\label{fig_bp}
\end{figure} 

\section{Gauge theory from isotropic ferromagnet} 
We consider a system of classical spins parameterised by a dimensionless vector field $\bm S(\bm r)=S\bm n(\bm r)$, where $\hbar S$ (with $\hbar$ the reduced Planck constant) is the molecular spin. We will restrict ourself to the exchange energy describing an isotropic ferromagnet and neglecting the dipolar and boundary terms. The exchange energy is explicitly ${\cal E}=\int({\rm d}V/2a) J(\nabla \bm S)^2$, with $J$ being the exchange energy constant, $a$ the lattice parameter and the integral extends over the whole volume. Assuming a given magnetic texture, $\bm n_0(\bm r)$, to be  a solution that minimizes the energy functional, we are interested in the fluctuations (SWs) around it. We consider the field $\bm n(\bm r,t)=\bm n_0(\bm r)+\bm \eta(\bm r,t)$, where $\bm\eta(\bm r,t)$ can be considered small and will encode the SWs dynamics. 
As fluctuations of a normalised vector field SWs are always in the plane normal to the magnetization. To take advantage of this constraint we use the local transformation $\Rb^{-1}$ that rotates the magnetic texture and put it in the $z$-axis as $\Rb^{-1}\bm n_0=\hat {\bm z}$. After the transformation, the fluctuations are in the normal plane to $z$ and can be written using a three-dimensional vector $\bm\omega=(\omega_1,\omega_2,\omega_3)$ defined by the relation $\Rb^{-1}\bm \eta=\hat {\bm z} \times \bm \omega=(-\omega_2,\omega_1,0)$. This relation shows us that, as a consequence of the normalization,  there are only two-components of $\bm \omega$ that counts for the SW dynamics and therefore it is enough to considering a two component field $\bm \omega\equiv(\omega_1,\omega_2)$. 
In this way we can express the normalised magnetization vector as $\bm n=\Rb\hat{\bm z}+ \Rb(\hat {\bm z} \times \bm \omega)$ and expand the exchange density $\epsilon=(JS^2/2a)(\nabla \bm n)^2$ until the second order in the fluctuations.  Using the fact that $\bm n_0$ is a local minimum of the functional and the unitarity of the rotation matrix, that is to say $\Rb^T\Rb=I$ we find the following SW Hamiltonian: 
\begin{equation}
\Eb=\frac{\hbar^2}{2m^*}((-i\vec{\nabla}+\vec{\Ab}) \bm \omega)^2-\bm\omega^T \Vb\bm\omega,
\label{eq_omega}
\end{equation}
where $m^*=\hbar^2/(4JSa^2)$ is the effective mass and the gauge (tensor) potential $\vec{\Ab}=- i \Rb_{\alpha 2}(\vec\nabla \Rb_{\alpha 1}) \tau$ (summation over repeated indices is assumed), with $\tau=-i\sigma_y$ where $\sigma_y$ is the corresponding Pauli matrix. In this way $\tau$ the generator of the group O(2). It is worth to note that in Eq. (\ref{eq_omega}) the square means the vector product and not the square of the complex norm. Explicitly the $\Vb=(\hbar^2/4m^*)\nu$ potential is
$$
\nu=\left(
  \begin{array}{ c c }
 (\Rb_{\alpha 2} \vec{\nabla} \Rb_{\alpha 1})^2-    (\vec{\nabla}\Rb_{\alpha 2})^2 &- \vec{\nabla} \Rb_{\alpha 1}\cdot\vec{\nabla} \Rb_{\alpha 2}\\
    - \vec{\nabla} \Rb_{\alpha 1}\cdot\vec{\nabla} \Rb_{\alpha 2} & (\Rb_{\alpha 2}\vec{\nabla} \Rb_{\alpha 1})^2-(\vec{\nabla}\Rb_{\alpha 1})^2
  \end{array} \right)
$$
We can exploit the symmetry of the problem writing the rotation matrix $\Rb$ in function of the Euler rotation angles $\Theta$, $\Phi$ and $\Psi$ of the original magnetic texture $\bm n_0=(\sin \Theta  \cos \Phi,\sin \Theta  \sin \Phi,\cos \Theta)$. In this representation both potentials take a simple form: 
\begin{equation}
\vec \Ab=i\bm A \tau,
\end{equation}
where explicitly, the gauge vector potential field $\bm A$ is:
\begin{equation}
\bm A=\cos \Theta \nabla \Phi+\nabla \Psi,
\label{vecA}
\end{equation}
and the $\Vb$ potential can be written as $\Vb=S^T V(\Theta,\Phi)S$, where $S=e^{\Psi\tau}$ is a rotation in the gauge coordinate $\Psi$ and the <<scalar>> potential is
\begin{equation}
V=\frac{\hbar^2}{4m^*}\left(
  \begin{array}{ c c }
    (\nabla\Theta)^2 &-\sin\Theta\:\nabla\Theta\cdot\nabla\Phi \\
-\sin\Theta\:\nabla\Theta\cdot\nabla\Phi    &\sin^2\Theta(\nabla\Phi)^2
  \end{array} \right).
\label{scalarV}
\end{equation}
In these equations the gauge symmetry is explicit, $\Psi$ is the gauge freedom coming from the fact that we can locally rotate $\bm n_0$ around itself without changing the physics.  This symmetry endows a conservation law given by the current: $\vec{j}=(\hbar/m^*)(i\vec{\nabla}-\vec{\Ab})\bm \omega^T \tau \bm\omega$. This current is the best definition of the spin wave current.

In a family of textures one can go farther along this analogy. If the equations $(\nabla\Theta)^2=\sin^2\Theta(\nabla\Phi)^2$ and $\nabla\Theta\cdot\nabla\Phi=0$ are satisfied, the potential (\ref{scalarV}) becomes a multiple of the identity. These identities hold for the case of a Bloch point that we are investigating. Interestingly, if that is the case, if we define $\Psi=\omega_1+i\omega_2$ we can write an equation for a unique complex field without anomalous terms as a Schr\"odinger-like functional:
\begin{equation}
\hat H_{\text{eff}}=\frac{\hbar^2}{2m^*}|(-i\nabla+\bm A)\Psi|^2- V|\Psi|^2
\label{Heff}
\end{equation}
where the gauge field $\bm A$ can be considered a vector potential  giving an effective magnetic field felt by  a quantum particle. 

The conditions over the magnetization field that lead to $V\propto I$ are stringent. Nevertheless there are many particular cases where this condition holds and it has been exploited in the literature before. \cite{Gonzalez-2010aa,Ivanov-2007} It is interesting to point out that this kind of analogy has been exploited even in fluid dynamics. \cite{Coste-1999} Put it in the present form it is easy to recognize in which cases we can write a Schr\"odinger equations for SWs.  
For the effective Hamiltonian Eq. (\ref{Heff}) there is a Noether conserved current done by $\bm j=(\hbar/m^*)[2\operatorname{Im}(\Psi^\dagger\nabla \Psi )- \bm A|\Psi|^2]$.

\section{Bloch point SWs and the Aharonov-Bohm effect} 
We can address now the problem of the scattering and SWs interference around a Bloch point (BP). The BP is characterised by a magnetization field that covers the whole sphere around the center of the soliton, which is, therefore, singular at that point for a normalised magnetization field. Its form, see Fig.(\ref{fig_bp}), is given by the field $\Theta(\bm r)=p\theta+\gamma$, $\Phi(\bm r)=q\phi$, where $(\theta,\phi)$ are the spherical coordinates of the space and $p$ and $q$ are integers that give the topological charge $Q=pq$ of the BP. \cite{Doring-1968bf}

Here we will choose for simplicity $p=1$, so $Q=q$. The conditions given a Schr\"odinger equations are satisfied for BPs and the effective Hamiltonian for the linear oscillations of SWs around the BP can be written as in the form of Eq. (\ref{Heff}), with $\mathbf{A}=\frac{q\cot\theta}{r}\hat\phi$ and $V(r)=-\frac{\hbar^2}{4m}\frac{ q^2}{r^2}$.  \cite{Elias-2014gf} In this way, around the BP, we have that the effective magnetic field is given by  a monopole-like magnetic field (known as Dirac monopole \cite{dirac-1931aa}) generated by a <<magnetic charge>> $q$. The vector potential is singular on a line whose location depends on the chosen gauge. In the present gauge we have $\nabla \cdot \bm A=0$, the so-called Coulomb gauge, the singularity line is along the whole $z$ axis and the potential is symmetric, see Fig. (\ref{fig_bp}), this is known as the Schwinger's potential. \cite{Schwinger1976451,Shnir-2005uq}
 
We can use the fact that when SW propagates the wave function acquires a magnetic phase.  \cite{Aharanov-1959} The phase difference between two interfering arbitrary paths can be calculated considering the integral of the potential vector:
\begin{equation}
\Delta \phi_{AB}=\oint\bm A\cdot \bm{ds},
\label{eq_ABphase}
\end{equation}
where $\bm{ds}$ is an infinitesimal part of the trajectory. From Stoke's theorem the expression in Eq. (\ref{eq_ABphase}) is the magnetic flux through the surface defined by the closed loop. The two paths can be continuously deformed in order to enclose a solid angle $\Omega$ of a spherical surface. Thus, the magnetic flux enclosed by the two paths can be given by $\Delta\Phi=\int qd\Omega=q\phi\int \sin\theta d\theta$, where $\theta$ is the polar angle.
 It has been shown \cite{Wu-1976,Lipkin-1982,Wilczek-1982,Fukuhara-1983,beche-2014} that this quantity is directly related with the magnetic charge, giving
\begin{equation}
\Delta \phi_{AB}=2q\phi,
\end{equation}
which is equivalent in our case to the statement that there is an Aharonov-Bohm (AB) phase directly proportional to the topological charge of the BP. 
 
\begin{figure}
\centering
\includegraphics[scale=0.3]{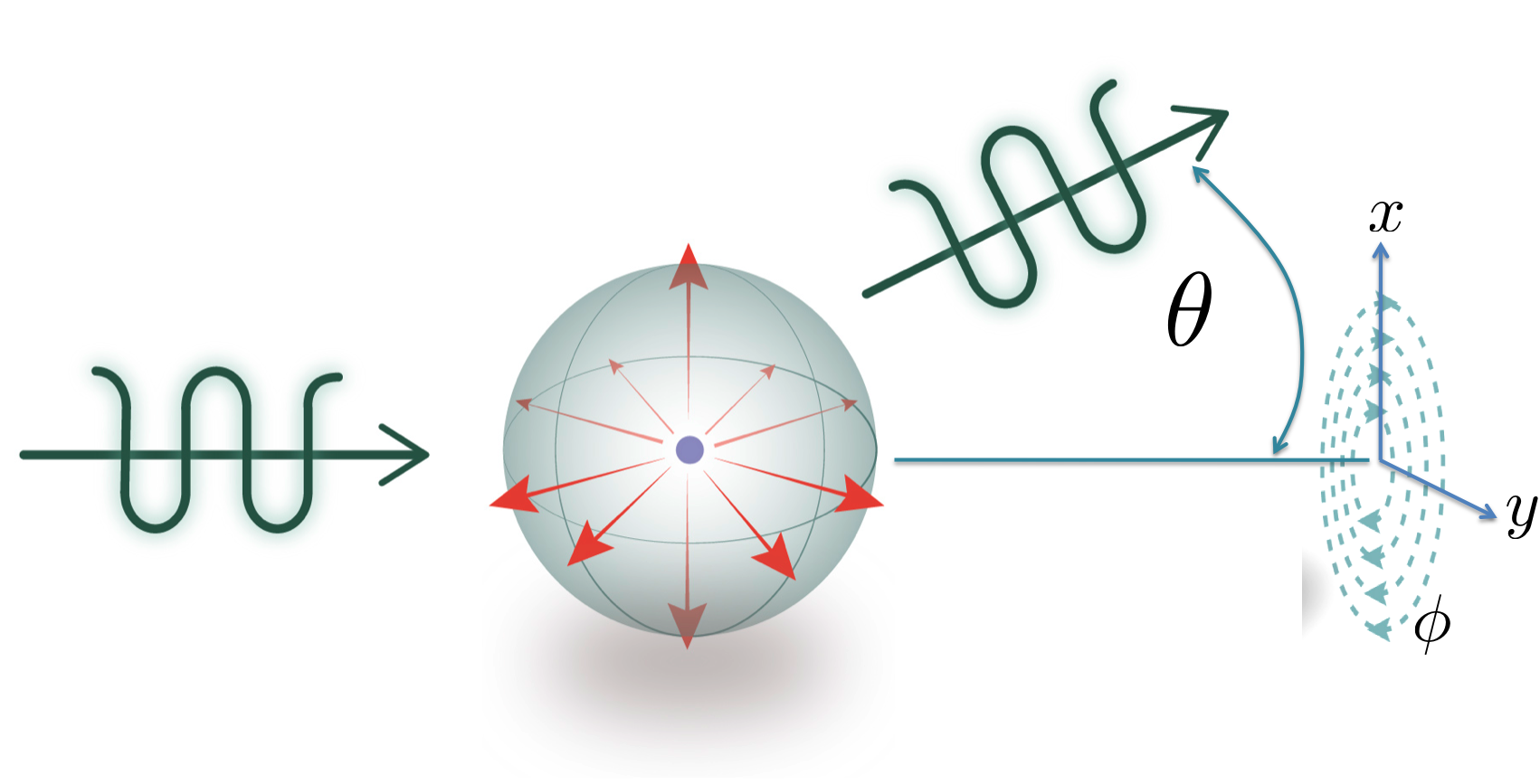}
\caption{Scattering of a spin wave by a magnetic Bloch point. The problem is equivalent to the scattering of a charged object by a magnetic monopole. In both cases the outgoing wave has the form of a vortex.}
\label{fig_bp}
\end{figure} 
The BP field leads to an integer $m$ ($q = m/2$), resulting in a perfect phase vortex of topological charge $m$. Thus, the interaction of a plane SW with a BP enables the typical azimuthal AB phase shift to occur and vortex SW states to be created. The Noether current can be readily calculated $\bm j_\phi=-(\hbar/m^*)|\Psi|^2(q(1-\cos\theta)/(r\sin\theta))$.\\

The analytical solution to this problem is known from the theory of magnetic monopoles \cite{Schwinger1976451,Boulware-1976} and can be written as  $\Psi(r,\theta)=\sum_{l=q}^{\infty}\sum_{m=-l}^{l}F_l(r,\theta)e^{i m\phi}$, where $F_l(r,\theta)$ is a complicated function including the generalised spherical  harmonics and the spherical Bessel functions (see \cite{Elias-2014gf} and the citations there in)  and $m$ an integer in the shown interval. It is worth to note that $m$ is directly related with the topological charge of the BP as the sum begins with $l=q$, meaning that in a non-trivial topology ($q\neq 0$) there are always modes with $m\neq 0$. Just as in the monopole case, \cite{beche-2014} the SW wave function can be written as: \cite{Schwinger1976451}
\begin{eqnarray}
\Psi_{\text{out}}(\bm r)\sim\frac{1}{r}e^{ikr}f(\theta)e^{iq\phi}=\Psi(r,\theta)e^{im\phi},
\label{PhaseChangeWave}
\end{eqnarray}
known as a vortex state.

\section{Magnon holography} 
In order to test these results, an experiment using holography can be used. Holography is the technique in which we produce the interference between an arbitrary beam with a coherent one, which serves as reference. \cite{fowles-1968aa} The interference pattern allows us to obtain the whole information about the original beam. Magnon holography is based upon the same principle by using SWs and it has recently appeared as a convenient technique to study SWs and to the rise of new technologies for data processing. \cite{Khitun-2013} In the present case we take advantage of the known properties of the interference of SWs to propose a mechanism to detect BPs. To do so, let us suppose and incident plane wave (IW) of SWs has been prepared to propagate in the $z$ direction and to pass across the BP. The phase distribution on this plane changes according to Eq. (\ref{PhaseChangeWave}) when a BP is set in the middle of the path.
 This change could be recorded by using the hologram principle by making the incident wave (IW) interfere with a reference plane wave (RW) $\psi_R=A e^{i\bm k\cdot \bm r}$ that does not pass in the BP region and having the same momentum and SW amplitude. The RW propagates in a different direction, with an angle $\varphi$ with respect to the SW passing through the BP. Reconstruction of the IW in an optical-like form could be performed such as a hologram by using Brillouin light scattering. \cite{Azevedo-1992} The phase of the IW is also reproduced and could be visualised by means of the interference between this wave and another optical wave. 
Visualisation is in the form of a set of equal-phase lines, i.e., a contour map of the spatial distribution of the magnetostatic volume wave modes that is, the <<phase changes>> when the RW is adjusted to be the same as the object wave for an empty specimen (in the present case, a ferromagnetic state). Then, the comparison wave is tilted and as well as an electron interference pattern, \cite{beche-2014} we have an interferogram consisting of fringes. Thus, the magnon vortices could be seen by using an apparatus similar to that one proposed by Fukuhara \textit{et al.}\cite{Fukuhara-1983} in the context of electron scattering, associated to the experimental arrangement given in Ref. \cite{Azevedo-1992}, which uses the Brillouin scattering setup for the optical detection of the component of the magnetization. 

\begin{figure}
\centering
\includegraphics[scale=0.4]{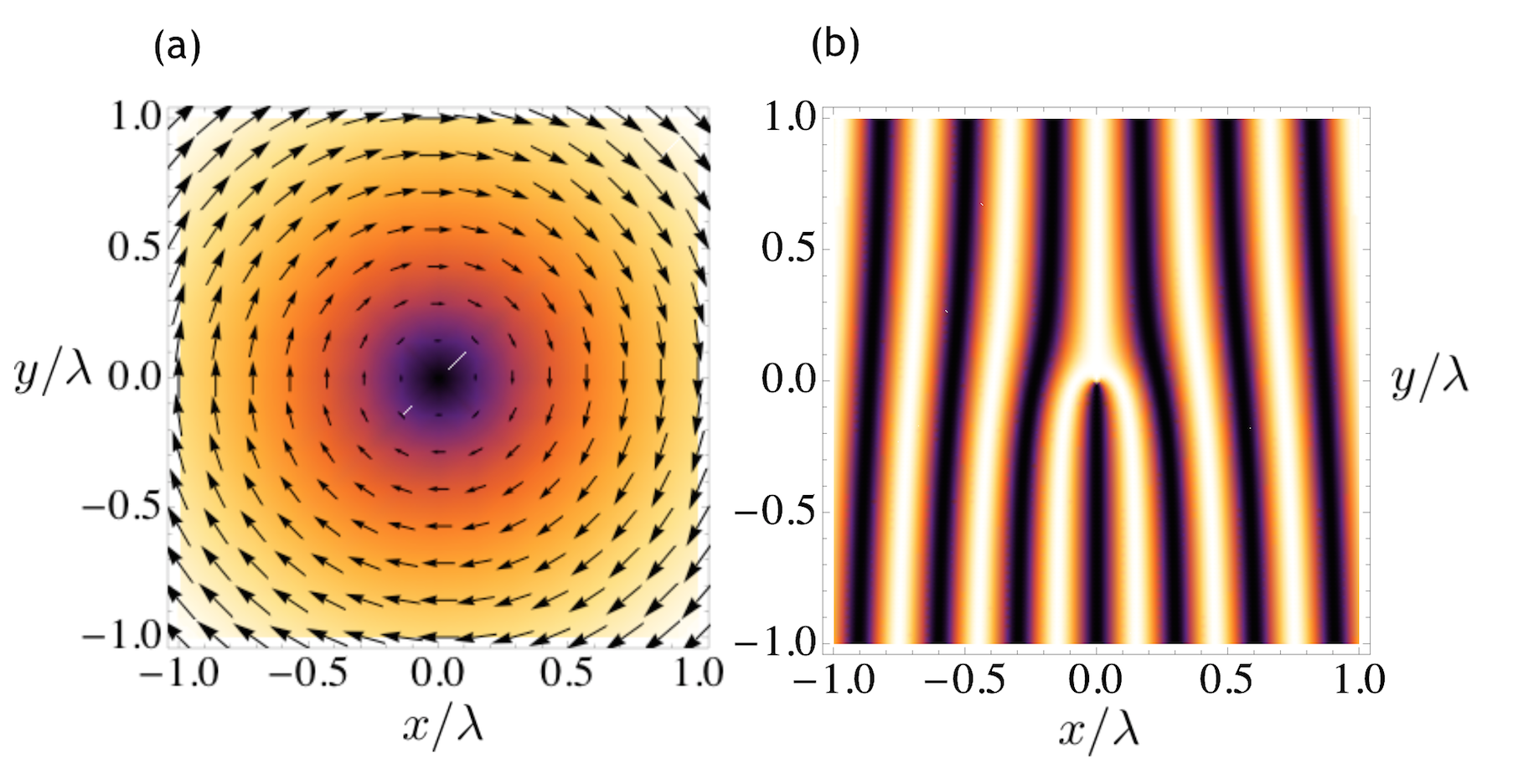}
\caption{(a) Magnon current associated with a magnon vortex. The plane of the vortex is perpendicular to the incoming wave. (b) Interference pattern between the magnon scattered by a BP for $q=1$ and  a free magnon forming with the former an angle $\varphi=\pi/2$. $\lambda$ is ten times the wavelength of the waves. }
\label{Experiment}
\end{figure}

Considering that the IW and the RW have the same amplitude and initial phases, and assuming that IW is perpendicularly incident to the recording plane, parameterised by the coordinates $(x,y)$, the intensity of interference pattern is given, in the proximity of the z-axis is: 
\begin{equation}
\frac{I}{2\mathcal{A}}=1+\cos(kx\sin\varphi-q\phi),
\end{equation}
where $\mathcal{A}$ is twice the square of the amplitude of the waves (that we considerer the same for both) and the direction of propagation of the IW is the axis $z$ and the k-vector of the RW lies, without lose of generality, in the $(x,z)$ plane (see Fig. \ref{fig_bp}). The interference patterns for different angles between the beams are shown in Fig. \ref{Experiment}. It can be noted that the interference pattern looks like a dislocation defect in a solid, showing the presence of a singularity in the system. Such singularity is associated with the string appearing in the potential vector. If the RW is going in the opposite direction of the IW ($\varphi=\pi$), the interference pattern is radial, showing the singularity in the center of the scattered SW vortex. The $\varphi$ angle between the two waves changes the scale of the pattern, but the dislocation remains, which makes the dislocation pattern robust.\\

\section{Conclusions}
In this paper we predict the appearance of a SW vortex for SWs propagating in the presence of a BP and we propose a way to detect using magnon holography. To study the SWs in the presence of a magnetic texture in a isotropic magnet we work in the local rotated frame where magnetization is in the ferromagnetic state. In this frame we were allowed to show that the effective theory for the SWs is a gauge theory with O(2) symmetry. We showed in which cases the texture support SWs that have a dynamics equivalent to that of a quantum particle one where the information about the texture is in the scalar and vector potential of the Schr\"odinger hamiltonian. For BPs the non-trivial topology of the texture gives a singular potential of a magnetic monopole and the azimuthal part of the scattered SW reflects the singularity by developing a vortex state. This can be indirectly detected by the interference of a free SW with the same wave vector, and the interference pattern has a robust and recognisable structure. This measuring process can be applied to other SWs in order to detect and control magnetic solitons.

\section{Acknowledgments}
R.G.E thanks Conicyt Pai/Concurso Nacional de Apoyo al Retorno de Investigadores/as desde el Extranjero Folio 821320024. V.L.C.S. thanks the Brazilian agency CNPq (Grant No. 229053/2013-0),
for financial support. The authors acknowledge funding from Proyecto Fondecyt No. 1150072, Proyecto Basal FB0807-CEDENNA, and by Anillo de Ciencia y Tecnonolog\'ia ACT 1117.

\bibliography{SWV}
\bibliographystyle{abbrv}

\end{document}